\documentstyle[prd,aps,preprint,epsfig,tighten]{revtex}
\begin{document}
\draft
\title{Solar neutrino event spectra:\\
	Tuning SNO to equalize Super-Kamiokande }
\author{G.L.\ Fogli, E.\ Lisi, A.\ Palazzo  }
\address{Dipartimento di Fisica and Sezione INFN di Bari,\\
                   Via Amendola 173, I-70126 Bari, Italy}
\author{F.L.\ Villante}
\address{Dipartimento di Fisica and Sezione INFN di Ferrara,\\
                   Via del Paradiso 12, I-44100 Ferrara, Italy}
\maketitle
\begin{abstract}
The Super-Kamiokande (SK) and the Sudbury Neutrino Observatory (SNO)
experiments are monitoring the flux of $^8$B solar neutrinos through the 
electron energy spectrum from the reactions $\nu_{e,\mu,\tau}+e^-\to
\nu_{e,\mu,\tau}+e^-$ and $\nu_e+d\to p+p+e^-$, respectively. We show that  the
SK detector response to $^8$B neutrinos in each bin of the electron energy
spectrum (above 8 MeV) can be approximated,  with good accuracy, by the SNO
detector response in an appropriate electron energy range (above 5.1 MeV). For
instance, the SK response in the bin  $[10,\,10.5]$ MeV is reproduced
(``equalized'') within $\sim 2\%$ by the SNO response in the range
$[7.1,\,11.75]$ MeV. As a consequence, in the presence of active neutrino
oscillations,  the SK and SNO event rates in the corresponding energy ranges
turn out to be linearly related, for any functional  form of the oscillation
probability. Such equalization is not spoiled by the possible contribution of
{\em hep\/} neutrinos (within current phenomenological limits).  In
perspective, when the SK and the SNO spectra will both be measured with high
accuracy,  the SK-SNO equalization can be used to determine the {\em
absolute\/} $^8$B neutrino flux, and to cross-check the (non)observation of
spectral deviations in SK {\em and\/} SNO. At present, as an exercise, we use
the equalization to ``predict'' the SNO energy spectrum, on the basis of the
current SK data. Finally, we briefly discuss some modifications or limitations
of our results in the case of sterile $\nu$ oscillations and of relatively
large Earth matter effects.
\end{abstract} 

\section{Introduction}

The origin of the deficit of solar neutrino events 
\cite{Cl98,Fu96,Ab99,Ha99,GNOC,SK00} with respect to the standard solar model
(SSM) predictions \cite{NuAs,BP00,JNBH} is expected to be clarified by the
measurement of events induced by $^8$B neutrinos in the Super-Kamiokande (SK)
\cite{Su00} and Sudbury Neutrino Observatory (SNO)  \cite{SNOB} experiments.

The SK experiment has presented accurate, high-statistics measurements of the
energy spectrum of electrons \cite{Su00} produced by the elastic scattering
(ES) reaction 
\begin{equation}
 \nu_{e,a} + e^- \to \nu_{e,a} + e^-\ ,
\label{ES}
\end{equation}
which can proceed either through $\nu_e$ or, in the presence of flavor
oscillations, also through  the other two active neutrinos $\nu_a$
$(a={\mu,\,\tau})$, with cross sections $\sigma^e$ and $\sigma^a$,
respectively.

The SNO experiment has presented preliminary results of the energy spectrum of
electrons \cite{SNOB}  produced by the charged-current (CC) absorption reaction
\begin{equation}
 \nu_{e} +  d \to p + p + e^-\ ,
\label{CC}
\end{equation}
which can proceed only through $\nu_e$,  with  cross section $\sigma^c$.

Both in SK and in SNO, the neutrino interaction  process, as well as the
subsequent  electron detection, tends to degrade the original $\nu$-spectrum
information in the final $e$-spectrum, according to the following ``energy
flow'':
\begin{equation}
E_\nu{\rm\ spectrum}
\stackrel{\em int.}{\longrightarrow} 
E'_e{\rm\ spectrum}
\stackrel{\em det.}{\longrightarrow}
E_e{\rm\ spectrum}\ ,
\label{flow}
\end{equation}
where $E_\nu$ is the $\nu$ energy, $E'_e$ is the total (true) electron energy,
and $E_e$ is the  total (measured) electron energy. The latter two energies are
generally different, as a consequence of the finite detector energy resolution
\cite{BaLi,Faid,300d} associated to the photon counting statistics. As far as
one is concerned with  the electron event rates in a given energy range
$[E_e^{\min},\,E_e^{\max}]$, the above energy transfer can be completely
characterized in terms of the so-called {\em response function\/} $\rho(E_\nu)$
associated to such range  and to the $^8$B neutrino source \cite{Faid,Vill}.%
\footnote{The response function basically represents the (unoscillated)
spectrum of  {\em interacted} neutrinos, in contrast with the unobservable
spectrum of {\em emitted\/} solar neutrinos \protect\cite{Faid,Vill}.}

In this work, we analyze the SK and SNO response function  to $^8$B neutrinos
in specific energy ranges. In Sec.~II we present our basic result: for each
spectrum bin $[E_e^{\min},\,E_e^{\max}]$ in SK (above 8 MeV), we  find a
suitable energy range $[\tilde{E}_e^{\min},\,\tilde{E}_e^{\max}]$ in SNO (above
5.1 MeV)%
\footnote{In the context of this paper, it is useful to adopt a different
notation for the observed electron energy in SK $(E_e)$ and SNO 
$(\tilde{E}_e)$.}
where the corresponding SK and SNO response functions are equal within a few
percent. As a consequence of such approximate ``equalization'', we show in
Sec.~III that, in the presence of active neutrino oscillations, the SK and SNO
event rates in the corresponding energy ranges are linked by a one-to-one
(linear) relation,  independently of the functional form of the oscillation
probability. Moreover, such relation allows the determination of the absolute
$^8$B $\nu$ flux. In perspective, the SK-SNO spectral equalization will be
useful to cross-check possible (non)observations of spectral deviations in the
two experiments, as well as to determine the absolute $^8$B $\nu$ flux
independently of the neutral-current (NC) SNO measurement (to be performed in
the near future \cite{SNOB}).  At present, pending both NC data and official CC
data from SNO, we use in Sec.~III the SK-SNO relation to ``predict'' the CC
spectrum in SNO, on the basis of the current SK energy  spectrum. In Sec.~IV we
show that the previous results are not spoiled by the possible contribution of
{\em hep\/} neutrinos  (within present phenomenological limits). Finally, we
discuss in Sec.~V some modifications or limitations of our results  in the
presence of sterile $\nu$ oscillations and of relatively large Earth matter
effects. Conclusions are presented in Sec.~VI.

The present work builds upon a previous paper \cite{Vill} in which a similar
strategy was devised to put in one-to-one relation the {\em total event
rates\/} of SK and SNO, rather than their {\em spectral rates\/} in each energy
bin. The reader is referred to the bibliography in \cite{Vill} for earlier (but
less realistic) attempts to  find relations between SK and SNO (spectral) rates
in a model-independent way.

\section{SK and SNO RESPONSE FUNCTIONS}

\subsection{Definitions}

As mentioned in the Introduction, the response function $\rho(E_\nu)$
characterizes the interaction+detection process for a given solar $\nu$ source
($^8$B neutrinos in our case) and for a given electron energy interval
($[E_e^{\min},\,E_e^{\max}]$ in SK and
$[\tilde{E}_e^{\min},\,\tilde{E}_e^{\max}]$ in SNO). The three classes of
response functions  relevant for our work are
\begin{eqnarray}
\rho_B^e(E_\nu,\,[E_e^{\min},\,E_e^{\max}]) &=& {\rm SK\ }(\nu_e,\,e) 
{\rm\ ES\ response\ function}\ ,
\label{rhoBe}\\
\rho_B^a(E_\nu,\,[E_e^{\min},\,E_e^{\max}]) &=& {\rm SK\ }(\nu_a,\,e) 
{\rm\ ES\ response\ function}\ \ (a=\mu,\tau)\ ,
\label{rhoBa}\\
\rho_B^c(E_\nu,\,[\tilde{E}_e^{\min},\,\tilde{E}_e^{\max}]) 
&=& {\rm SNO\ }(\nu_e,\,d) 
{\rm\ CC\ response\ function}
\label{rhoBc}\ .
\end{eqnarray}

Such functions are defined in terms of the standard $^8$B energy spectrum
$\lambda_B(E_\nu)$ \cite{Bspe}, of the differential cross sections
($d\sigma^e/dE'_e$ and $d\sigma^a/dE'e$ for elastic scattering  \cite{EScs} and
$d\sigma^c/d\tilde{E}'_e$ for CC absorption \cite{CCcs}), and of the detector
resolution functions ($R_{\rm SK}$ \cite{SKre}  and $R_{\rm SNO}$ \cite{SNre}):
\begin{eqnarray}
\rho_B^e &=& \frac{\displaystyle
\lambda_B(E_\nu)\int_{E_e^{\min}}^{E_e^{\max}}dE_e
\int_0^{E_\nu}dE'_e\,
\frac{d\sigma^e(E_\nu,\,E'_e)}{dE'_e}\,R_{\rm SK}(E_e,\,E'_e)}
{\sigma^e_B[E_e^{\min},\,E_e^{\max}]}\ ,\label{rhoe}\\[4mm]
\rho_B^a &=& \frac{\displaystyle
\lambda_B(E_\nu)\int_{E_e^{\min}}^{E_e^{\max}}dE_e
\int_0^{E_\nu}dE'_e\,
\frac{d\sigma^a(E_\nu,\,E'_e)}{dE'_e}\,R_{\rm SK}(E_e,\,E'_e)}
{\sigma^a_B[E_e^{\min},\,E_e^{\max}]}\ ,\label{rhoa}\\[4mm]
 \rho_B^c &=& \frac{\displaystyle
\lambda_B(E_\nu)\int_{\tilde{E}_e^{\min}}^{\tilde{E}_e^{\max}}d\tilde{E}_e
\int_0^{E_\nu}d\tilde{E}'_e\,
\frac{d\sigma^c(\tilde{E}_\nu,\,\tilde{E}'_e)}
{d\tilde{E}'_e}\,R_{\rm SNO}(\tilde{E}_e,\,\tilde{E}'_e)}
{\sigma^c_B[\tilde{E}_e^{\min},\,\tilde{E}_e^{\max}]}\label{rhoc}\ ,
\end{eqnarray}
where the denominators $\sigma_B^{e,a,c}$ represent the $^8$B neutrino cross
sections for producing an electron with observed energy in the specified range,
as obtained by integrating over $E_\nu$ the corresponding numerators in
Eqs.~(\ref{rhoe})--(\ref{rhoc}).%
\footnote{It follows that the functions $\rho^{e,a,c}_B(E_\nu)$ in
Eqs.~(\protect\ref{rhoe})--(\protect\ref{rhoc}) are normalized to unity.}

Throughout this paper, the response functions are calculated through numerical
integration of the above expressions,  Eqs.~(\ref{rhoe})--(\ref{rhoc}).
Alternatively, the response functions can be  obtained through MonteCarlo (MC)
simulations. In particular, the function $\rho_B^e$ is just the normalized  
$E_\nu$ spectrum (histogram) of the $^8$B neutrinos originating a simulated
electron with a measured energy $E_e\in[E_e^{\min},\,E_e^{\max}]$ in the SK
detector, in the absence of oscillations;  analogously for $\rho_B^c$ in SNO.
The function $\rho_B^a$ can be obtained in the same way as $\rho_B^e$, provided
that the differential cross section  $d\sigma^e/dE'_e$ is replaced  by
$d\sigma^a/dE'_e$ in the MC simulation. Eventually, if our results were adopted
by the SK and SNO collaborations in a joint SK-SNO analysis, the  corresponding
energy ranges and  response functions should be optimally calculated and
compared through MC simulations, so as to include also those minor detector
features (small energy variations of efficiencies, slightly nongaussian
resolutions, etc.) which require the knowledge of experimental details beyond
the scope of this paper.

\subsection{Correspondence between SK and SNO energy ranges}

In \cite{Vill} it has been shown that the SK and SNO response functions for the
whole spectrum can be equalized, to a good approximation, by an appropriate
choice of the corresponding  electron energy thresholds. In this work we make
an important further step, by showing that the SK response function in each
bin  $[E_e^{\min},\,E_e^{\max}]$ (above 8 MeV)  is approximately equalized by
the SNO response function in a suitably chosen range
$[\tilde{E}_e^{\min},\,\tilde{E}_e^{\max}]$,
\begin{equation}
\rho^{e,a}_B(E_\nu,[E_e^{\min},\,E_e^{\max}]) \simeq
\rho^{c}_B(E_\nu,[\tilde{E}_e^{\min},\,\tilde{E}_e^{\max}]) \ .
\label{basic}
\end{equation}
In this way one can extend  the SK-SNO correspondence from {\em total rates\/}
\cite{Vill} to {\em energy spectra\/}.

Technically, for each SK bin $[E_e^{\min},\,E_e^{\max}]$, we have determined
the extrema of the SNO range  $[\tilde{E}_e^{\min},\,\tilde{E}_e^{\max}]$ by
requiring minimization of the integral difference
\begin{equation}
\Delta = \int dE_\nu\, |\rho_B^e(E_\nu)-\rho_B^c(E_\nu)|\ ,
\label{Delta}
\end{equation}
which would be zero if Eq.~(\ref{basic}) were exactly satisfied. The results
are given in Table~I and in Figs.~1 and 2, as we now discuss.

Table~I gives, for each SK spectrum bin  above 8 MeV (labelled by a sequential
number $i=1,2,\dots,13$), the corresponding energy range
$[\tilde{E}_e^{\min},\,\tilde{E}_e^{\max}]$ in SNO (third column) where the
SK-SNO response function difference $\Delta$ (fourth column) is minimized. In
addition, the fifth and sixth column present the $^8$B neutrino cross sections
$\sigma_B^e$ and $\sigma_B^a$ $(a=\mu,\,\tau)$ for electron production in the
SK range $[E_e^{\min},\,E_e^{\max}]$ (together with their ratio in the seventh
column), while the last column presents the $^8$B neutrino cross sections
$\sigma_B^c$  for CC electron production in the corresponding SNO range
$[\tilde{E}_e^{\min},\,\tilde{E}_e^{\max}]$, as defined in 
Eqs.~(\ref{rhoe})--(\ref{rhoc}) and related comments. Figure~1 shows
graphically the SK-SNO energy range correspondence expressed by the first three
columns of Table~I.%
\footnote{The SK energy bins below 8 MeV are not considered in Fig.~1 and in
Table~I, since the corresponding SNO energy ranges would be (partly) below the
expected SNO threshold \protect\cite{SNOB} of $\sim 5$ MeV.} 

Figure~2 shows the approximate equality between the SK response functions
$\rho_B^{e,a}$ (solid curves) and the corresponding SNO response functions
$\rho_B^c$ (dotted curves) in a representative  subset of  energy ranges. In
the worst case ($i=13$, highest energy range) the integral difference $\Delta$
does not exceed $10\%$, and it is often much smaller in other ranges. It can be
seen, for instance, that $\rho_B^e$ for the interval  $E_e\in[10,\,10.5]$ MeV
in SK ($i=5$) is almost coincident with $\rho_B^c$ for the interval 
$\tilde{E}_e\in[7.1,\,11.75]$, $\Delta$ being as small as $2.1\%$ (see
Table~I).%
\footnote{Notice that the values of $\Delta_i$ in Table~I are calculated with
the input ingredients described in Sec.~II.A. Should such ingredients (e.g.,
the SNO resolution function) change, the extrema of the SNO ranges 
$[\tilde{E}^{\min},\tilde{E}^{\max}]_i$ minimizing $\Delta_i$  (and the values
of $\Delta_i$ themselves) should be recalculated for any $i$-th interval.}
In addition, the functions $\rho_B^{e}$ and $\rho_B^a$ in Fig.~1 are
graphically indistinguishable (they practically coincide with one and the same
solid curve for any range).

We can summarize such results by stating that there is a set $(i=1,2,\dots,13)$
of SK and SNO ranges ($[E_e^{\min},\,E_e^{\max}]$ and
$[\tilde{E}_e^{\min},\,\tilde{E}_e^{\max}]$, respectively) where the two
experiments are characterized by very similar responses to $^8$B neutrinos,
\begin{eqnarray}
\rho^{e}_B(E_\nu,[E_e^{\min},\,E_e^{\max}]_i) &=& 
\rho^{a}_B(E_\nu,[E_e^{\min},\,E_e^{\max}]_i) \ ,
\label{trivial}\\
\rho^{e}_B(E_\nu,[E_e^{\min},\,E_e^{\max}]_i) &\simeq& 
\rho^{c}_B(E_\nu,[\tilde{E}_e^{\min},\,\tilde{E}_e^{\max}]_i)
\label{nontrivial}\ .
\end{eqnarray}

A few comments are in order. The response function shape is governed by the
electron energy window $[E_e^{\min},\,E_e^{\max}]$ and by three functions: the
$\nu$ source spectrum $\lambda_B$, the normalized differential cross section
$\sigma_B^{-1}d\sigma_B/dE'_e$, and the detector resolution  $R(E'_e,E_e)$ [see
Eqs.~(\ref{rhoe})-(\ref{rhoc})]. Both in SK and in SNO, the bell-shaped
functions $\lambda_B$ and $R$ render the response functions also bell-shaped,
independently of the functional form of the differential cross section. Such
functional form, together with the chosen electron energy window, affects
instead the ``width at half maximum'' of the response function. In SK, the
normalized  cross sections as a function of the electron energy,
$(\sigma^e_B)^{-1}d\sigma^e_B/dE'_e$  and $(\sigma^a_B)^{-1}d\sigma^a_B/dE'_e$,
are  both very similar  for any fixed neutrino energy $E_\nu$ \cite{EScs}.
Therefore, it is not surprising that Eq.~(\ref{trivial}) holds with very high
accuracy.

Conversely, the approximate equality (\ref{nontrivial}) is nontrivial. In fact,
the  shape of the normalized CC cross section in SNO  is very different from
the ES case, being peaked at a total electron energy only slight smaller (by
$\sim 1.5$ MeV) than the  $\nu$ energy, with a long tail at low $\tilde{E}'_e$
\cite{CCcs}. Therefore, if  {\em identical\/} electron energy ranges were taken
for SK and SNO, 
$[E_e^{\min},\,E_e^{\max}]=[\tilde{E}_e^{\min},\,\tilde{E}_e^{\max}]$, the SNO
response function would be much narrower  (and peaked to slightly higher
neutrino energies) than the SK response function (e.g., compare Figs.~2 and 7
in \cite{Faid}). The functional difference in the cross sections and in their
$Q$-values (and also, to same extent, the difference in the detectors
resolution, $R_{\rm SK}\neq R_{\rm SNO}$), render the SNO response functions
{\em tipically different\/}  from those of SK. In order to achieve an {\em
approximate equalization}, as the one shown in Fig.~2, one needs 
$[E_e^{\min},\,E_e^{\max}]\neq [\tilde{E}_e^{\min},\,\tilde{E}_e^{\max}]$ and,
in particular: (i) The SNO range width  
$\tilde{E}_e^{\max}-\tilde{E}_e^{\min}$ must be larger than the SK range width 
$E_e^{\max}-E_e^{\min}$ (so as to ``broaden'' the SNO response function);  and
(ii) The central value of the SNO range must be slightly lower than for SK (so
as to tune  the SNO response function peak at the same energy as in SK). 
Indeed, both conditions $(i)$ and $(ii)$  are fulfilled by the correlated
SK-SNO energy ranges reported in Table~I.%
\footnote{The broadening of the SNO energy ranges with respect to SK implies
that the intervals $[\tilde{E}^{\min},\,\tilde{E}^{\max}]_i$ are partly 
overlapping, as also evident from Fig.~1.} 
We will discuss in the next section how to exploit the approximate equalization
of the SK and SNO response functions in such correlated energy ranges.

\section{Linking SK and SNO spectral rates for active $\nu$ oscillations}

In this section we derive a linear relation between the SK and SNO event rates,
valid for any of the thirteen couples of SK-SNO energy ranges reported in
Table~I, and for any functional form of the {\em active\/} neutrino oscillation
probability. We start with the definition of the unoscillated rates, to which
the experimental rates are usually normalized.

The standard $^8$B neutrino event rate (per target electron,  in each $i$-th
bin of the SK electron spectrum),  in the absence of oscillations, is given by
\begin{equation}
R^{0,i}_{\rm SK}=\Phi^0_B\,\sigma_B^{e,i}\ ,
\label{R0_SK}
\end{equation}
where the cross section $\sigma_B^{e,i}$ is given in the 5th column of Table~I,
while $\Phi_B^0$ is the $^8$B neutrino flux from the reference standard solar
model ``BP 2000'' \cite{BP00},  $\Phi^0_B=5.15\times 10^6$ cm$^{-2}$s$^{-1}$.
Analogously, the standard rate in the corresponding $i$-th energy range of the
SNO spectrum is given by
\begin{equation}
R^{0,i}_{\rm SNO}=\Phi^0_B\,\sigma_B^{c,i}\ ,
\label{R0_SNO}
\end{equation}
where our evaluation of the cross section $\sigma_B^{c,i}$ is given in the last
column of Table~I.

In general, the event rates measured in SK and SNO can be different from the
previous expectations, as a result of either neutrino oscillations or of
deviations from the standard model predictions (or both). Here we assume {\em
active\/} neutrino oscillations, described by a generic $\nu_e$ survival
probability function $P_{ee}(E_\nu)\leq 1$. Deviations from the SSM flux are
parametrized through an unknown factor $f_B$ multiplying the standard flux
$\Phi_B^0$. The event rates measured in SK and SNO should then be equal to
\begin{eqnarray}
R^{i}_{\rm SK}&=& R^{e,i}_{\rm SK} + R^{a,i}_{\rm SK}\\
&=& f_B\,\Phi^0_B\,
\left[\sigma_B^{e,i}\,\langle P_{ee}\rangle^{e,i}_B
+\sigma_B^{a,i}\,\left(1-\langle P_{ee}\rangle^{a,i}_B\right)\right]\ ,
\label{R_SK}
\end{eqnarray}
and
\begin{equation}
R^{i}_{\rm SNO}=f_B\,\Phi^0_B\,\sigma_B^{c,i}\,\langle P_{ee}\rangle^{c,i}_B\ ,
\label{R_SNO}
\end{equation}
respectively.%
\footnote{In Eq.~(\protect\ref{R_SK}), the contributions from $\nu_e$ and
$\nu_a$ ($a=\mu,\tau$) are explicitly separated.} 
The terms $\langle P_{ee}\rangle^{X,i}_B$ ($X=e,\,a,\,c$) represent the average
survival probability in the $i$-th bin, weighted by the appropriate $^8$B
response function,
\begin{equation}
\langle P_{ee}\rangle_B^{X,i}=\int dE_\nu\,
\rho^{X,i}_{B}(E_\nu)\, P_{ee}(E_\nu)\ \ (X=e,\,a,\,c)\ .
\label{Pee}
\end{equation}
Notice that Eqs~(\ref{R_SK})--(\ref{Pee}) are exact, i.e., they are definitions
which hold without any approximation.

We now apply the approximate equality of the SK and SNO $^8$B response
functions expressed by Eqs.~(\ref{trivial},\ref{nontrivial}). Such equality
implies that the average probabilities in (\ref{Pee}) are also approximately
equal to each other,
\begin{equation}
\langle P_{ee}\rangle^{i}_B \equiv
\langle P_{ee}\rangle^{e,i}_B =
\langle P_{ee}\rangle^{a,i}_B \simeq 
\langle P_{ee}\rangle^{c,i}_B \ \ (i=1,\dots,13)\ ,
\label{Pequal}
\end{equation}
within errors smaller than the corresponding values of $\Delta$
[Eq.~(\ref{Delta})] given in Table~I:
\begin{equation}
\delta \langle P_{ee}\rangle_B^i \equiv \left| 
\int dE_\nu \, P_{ee}(E_\nu)\, [\rho^{e,i}(E_\nu)_B-\rho_B^{c,i}(E_\nu)]
\right|
\leq
\int dE_\nu \, P_{ee}\, \left| 
\rho^{e,i}_B-\rho_B^{c,i}\right| \leq \Delta_i\ .
\label{error}
\end{equation}
It turns out that the error $\delta \langle P_{ee}\rangle_B^i$ is often much
smaller than its upper limit $\Delta_i$ for typical oscillation solutions of 
the solar neutrino problem, being usually of $O(1\%)$.%
\footnote{The error $\delta \langle P_{ee}\rangle$ is smaller than 
$\Delta_i$ even when $P_{ee}$ is largest
[$P_{ee}(E_\nu)\equiv1$, no oscillations] since
the difference $\rho^{e,i}(E_\nu)_B-\rho_B^{c,i}(E_\nu)$ changes sign
as a function of $E_\nu$ (Fig.~2).
See also the related discussion in Sec.~V of \protect\cite{Vill}.}

In other words, Eqs.~(\ref{Pequal},\ref{error}) show that, for any $i$-th
couple of SK and SNO energy intervals in Table~I, the two experiments probe
(within a typical accuracy of a percent) one and the same average survival
probability $\langle P_{ee}\rangle^i_B$ for $^8$B neutrinos. Therefore,  the
normalized SK and SNO $i$-th event rates can be written as
\begin{eqnarray}
r^i_{\rm SK} &\equiv& \frac{R^i_{\rm SK}}{R^{0,i}_{\rm SK}} =
f_B\,\left[
 \langle P_{ee}\rangle^i_B + \frac{\sigma^{a,i}_B}{\sigma^{e,i}_B}
(1-\langle P_{ee}\rangle^i_B) \right]\ ,
\label{r_SK}\\
r^i_{\rm SNO} &\equiv& \frac{R^i_{\rm SNO}}{R^{0,i}_{\rm SNO}} =
f_B\,
 \langle P_{ee}\rangle^i_B \ .
\label{r_SNO}
\end{eqnarray}
An immediate consequence is that the presence of NC events in the $i$-th SK
bin, due to $\nu_a$ interactions ($a=\mu,\tau$), can emerge in a
model-independent way by observing $r^i_{\rm SK}-r^i_{\rm SNO}>0$:
\begin{equation}
r^i_{\rm SK}-r^i_{\rm SNO}>0 \ \Longrightarrow\ \nu_e\to\nu_{\mu,\tau} {\rm\
channel\ open}\ .
\label{NC}
\end{equation}

By eliminating $\langle P_{ee}\rangle^i_B$  from Eqs.~(\ref{r_SK},\ref{r_SNO}),
a linear relation between the SK and SNO normalized rates emerges in each
$i$-th couple of energy intervals $(i=1,\dots,13)$:
\begin{equation}
r^i_{\rm SK} = r^i_{\rm SNO} \left( 1-\frac{\sigma^{a,i}_B}{\sigma^{e,i}_B}
\right) + f_B\,\frac{\sigma^{a,i}_B}{\sigma^{e,i}_B}\ \ 
\ ,
\label{main}
\end{equation}
where the numerical values  of $\sigma^{a,i}_B/\sigma^{e,i}_B$ are given in the
7th column of Table~I. Notice that, since the above equation does not depend on
the average probability $\langle P_{ee}\rangle_B^{i}$, it holds for any
functional form of the unaveraged survival probability  $P_{ee}(E_\nu)$, i.e.,
for any {\em active\/} $\nu$  oscillation solution of the  solar neutrino
problem.

For the benefit of the reader, we have explicited Eq.~(\ref{main}) for each
energy range in Table~II. In all the equations listed in last column of such
Table, the numerical coefficients  are obviously very similar, since the ratio
$\sigma^{a,i}_B/\sigma^{e,i}_B$ depends weakly on energy. The nonobvious fact
is that such equations hold with good accuracy for any functional form of
$P_{ee}(E_\nu)$, provided that the SNO energy ranges are chosen according the
third column of Table~II, for any fixed SK energy bin in the second column of
the same  table.

Equation~(\ref{main}), explicited in Table~II, represents the main result of
our work.  It implies a one-to-one correspondence between thirteen couples of
SK and SNO electron event rates induced by $^8$B neutrinos,  independently of
the specific {\em active\/} $\nu$  oscillation solution. It also explicitly
incorporates  possible deviations from the standard $^8$B flux through the
factor $f_B$. The physical content of such result can be expressed as follows:
for any $i$-th SK bin above 8 MeV, it turns out that there is a SNO energy
interval where the average suppression ($\langle P_{ee}\rangle_B^{i}$) of the
$\nu_e$ flux due to oscillations is the same in the two experiments. Then,
modulo factors as $f_B$ and $\sigma^{a,i}_B/\sigma^{e,i}_B$, one can get also
the average suppression of the $\nu_a$ flux in SK ($a=\mu,\tau$), and thus a
link between the SK and SNO rates which is independent of the specific
probability function $P_{ee}(E_\nu)$. We discuss now some possible applications
of Eq.~(\ref{main}).

When official CC spectrum data (and thus the $r^i_{\rm SNO}$'s)  will be
released by SNO, Eq.~(\ref{main}) will provide a determination of $f_B$,
namely, of the absolute $^8$B flux at the Sun ($f_B\times \Phi_B^0$). This fact
was derived in \cite{Vill} by using {\em total\/} SK and SNO rates with
appropriate thresholds; our  Table~II generalizes such result to several energy
spectrum ranges $(i=1,\dots,13)$,  so that the value of $f_B$ will be {\em
overconstrained\/}. Notice that such constraints on $f_B$ will {\em precede\/}
the independent determination of $f_B$ through the neutral current (NC) 
measurement planned in the near future in SNO \cite{SNOB}.

Table~II can also be used to cross-check possible spectral deviations (or their
absence) in SK and SNO. If there is (not) a specific spectral distortion
pattern  $\left\{r^i_{\rm SK}\neq {\rm const}\right\}_{1\leq i\leq 13}$ in the
SK spectrum above 8 MeV, a similar pattern must show up, independently of
$P_{ee}$, in the sequence $\left\{r^i_{\rm SNO}\right\}_{1\leq i \leq 13}$
above 5.1 MeV in SNO (within uncertainties), according to Table~II.  Although
it is intuitively clear that, if SK finds hints of spectral deviations, they
should also be found by SNO \cite{Hint},  Table~II provides a  quantitative and
well-defined way to make such cross-check, with the great advantage that no
prior assumption is needed about the (unknown and unobservable) function
$P_{ee}(E_\nu)$.

Pending official and definite CC spectrum data from SNO,%
\footnote{Preliminary SNO results have been reported in \protect\cite{SNOB}.}
we use Table~II to make a sort of ``prediction'' for the SNO spectrum, based on
the measured SK spectrum \cite{SKSP}.  Figure~3 shows, in the upper panel, the
normalized SK spectrum $r^i_{\rm SK}$ \cite{SKSP} above 8 MeV.  In each of the
thirteen bins $[E^{\min},\,E^{\max}]_i$, the height of the grey box is equal to
the $\pm 1\sigma$  statistical error $\delta r^i_{\rm SK}$. Inverting the
relations in Table~II, and assuming $f_B=1$, we propagate the SK data set
$\{r^i_{\rm SK}\pm \delta r^i_{\rm SK}\}_i$ to  a set of ``predicted rates''
$\{r^i_{\rm SNO}\pm \delta r^i_{\rm SNO}\}_i$ for SNO,  in the corresponding
energy intervals $[\tilde{E}^{\min},\,\tilde{E}^{\max}]_i$. The results are
shown in the lower panel of Fig.~3, where the grey boxes have now a width 
determined by $[\tilde{E}^{\min},\,\tilde{E}^{\max}]_i$, and are thus partially
overlapping. The meaning of such exercise is the following: In the presence of
generic active $\nu$ oscillations, and for $f_B=1$, the measured SNO spectrum
has to be consistent with the lower panel of Fig.~3, in order to be compatible
with the present SK data and their $\pm 1\sigma$~(stat.) errors. The exercise
in Fig.~3 can be repeated for $f_B\neq 1$ (not shown).

A final remark is in order, concerning the uncertainties affecting
Eq.~(\ref{main}). Besides the experimental uncertainties in the SK and SNO
rates, the main theoretical uncertainty is related to the {\em absolute\/} 
normalization of the $\sigma^c$ cross section for CC interactions in SNO, which
enters in the calculation of $R^{0,i}_{\rm SNO}$ [Eq.~(\ref{R0_SNO})] and thus
propagates to the normalized rate $r^i_{\rm SNO}$  [Eq.~(\ref{r_SNO})]. Such
error is estimated in \cite{CCer} to be $\sim 6\%$  at $1\sigma$, by comparing
different calculations;  a more definite evaluation (and possibly a reduction)
would be highly desirable. Compared with such uncertainty, the  error  induced
by the approximate SK-SNO equalization [Eq.~(\ref{error})] is typically less
relevant, except perhaps in the worst case ($i=13$ energy range). Finally,
notice that Table~II involves thirteen couples of quantities, which have
correlated errors. Besides the obvious bin-by-bin correlation of detector
systematics in SK  (and, independently, in SNO), the $^8$B spectrum shape
uncertainty \cite{Bspe} represents a systematic error in {\em common\/} to SK
and SNO. Moreover, since the energy ranges
$[\tilde{E}^{\min},\,\tilde{E}^{\max}]_i$ partially overlap in SNO (Fig.~1),
the corresponding SNO rates $r^i_{\rm SNO}$ are also statistically correlated.
All such ``complications'' of a joint SK-SNO analysis can actually be handled
by standard statistical techniques (e.g., covariance matrices); however,
pending detailed data and official evaluations of uncertainties in SNO, we do
not furtherly explore this matter at present.

\section{Effect of {\em\lowercase{hep}} neutrinos}

In this section we show that our main result [Eq.~(\ref{main})] is basically
preserved in the presence of  a nonnegligible {\em hep\/} neutrino contribution
to the SK and SNO event rates, modulo a redefinition of the factor $f_B$.

The latest evaluation (BP 2000) of the standard  {\em hep\/} flux is
$\Phi^0_{\em hep}=9.3\times 10^{-7}$ cm$^{-2}$~s$^{-1}$ \cite{BP00}, with large
(unquoted) uncertainties. We parametrize such uncertainty by  introducing
(analogously to $f_B$) a free parameter $f_{\em hep}$ multiplying $\Phi^0_{\em
hep}$. The neutrino energy spectrum at the Sun (as far as SK and SNO
are concerned) becomes then:
\begin{equation}
\Phi(E_\nu)=
f_B\, \Phi_B^0 \, \lambda_B(E_\nu)+
f_{\em hep}\, \Phi_{\em hep}^0 \, \lambda_{\em hep}(E_\nu)\ ,
\label{Flux}
\end{equation}
where $\lambda_{\em hep}(E_\nu)$ is the {\em hep\/} neutrino energy spectrum
\cite{JNBH}.

We recall that the slight ``excess'' of events in the high-energy tail of the
SK normalized spectrum \cite{Su00} is roughly consistent with the standard
contribution $(f_{\em hep}\sim 1)$ from {\em hep\/} neutrinos \cite{BP00}, and
provides a 90\% C.L.\ upper bound to such contribution,
$$
	f_{\em hep}\lesssim 3\ ,
$$
as derived in \cite{Su00} through an analysis with unconstrained $f_B$.  As an
example of relatively ``large'' {\em hep\/} flux we  consider then the
reference case $(f_B,\,f_{\em hep})=(1,\,3)$.

As for $^8$B neutrinos [Eqs.~(\ref{rhoBe})--(\ref{rhoBc})],  one can introduce
for {\em hep\/} neutrinos three new response functions,
\begin{equation}
\rho_{\em hep}^X(E_\nu) = {\rm response\ function\ to\ }{\em hep\ }{\rm
neutrinos\ }\ \ (X=e,\,a,\,c)\ ,
\label{rhohep}
\end{equation}
which are defined analogously to Eqs.~(\ref{rhoe})--(\ref{rhoc}), modulo the
replacement of $\lambda_B(E_\nu)$ with $\lambda_{\em hep}(E_\nu)$.
Correspondingly, one can define three new averaged probabilities,
\begin{equation}
\langle P_{ee}\rangle^X_{\em hep} = \int dE_\nu\,\rho^X_{\em hep}(E_\nu)
\,P_{ee}(E_\nu)\ \ (X=e,\,a,\,c)\ .
\end{equation}
Figure~4 shows the SK response functions $\rho_{\em hep}^e$ and $\rho_{\em
hep}^a$ (graphically coincident with the solid curve) in three representative
high-energy bins, together with the SNO response function $\rho_{\em hep}^c$
(dotted curve) in the corresponding energy ranges given in Table~I. The
response functions are somewhat different from each other, the difference
becoming rapidly larger in lower energy intervals (not shown). Therefore, we
cannot derive a reasonably approximate equality of the kind  $\langle
P_{ee}\rangle^e_{\em hep}\simeq \langle P_{ee}\rangle^c_{\em hep}$  [as it was
instead the case for $^8$B neutrinos, Eq.(\ref{Pequal})] However, this
potential problem practically disappears in the combination with $^8$B
neutrinos.

In fact, let us consider the combined B+{\em hep} response functions $\rho^X$
for a source flux as in Eq.~(\ref{Flux}). Such functions are given by
\begin{eqnarray}
\rho^X(E_\nu) &=& {\rm (B+{\em hep})\ response\ function}\ \ (X=e,\,a,\,c)
\\
&=& \frac{f_B\,\Phi^0_B\,\sigma^X_B\,\rho_B^X(E_\nu)+
f_{\em hep}\,\Phi^0_{\em hep}\,\sigma^X_{\em hep}\,\rho_{\em hep}^X(E_\nu)}
{f_B\,\Phi^0_B\,\sigma^X_B+
f_{\em hep}\,\Phi^0_{\em hep}\,\sigma^X_{\em hep}}
\\
&=&
\frac{\rho_B^X(E_\nu)+
\frac{f_{\em hep}\,\Phi_{\em hep}^0\,\sigma^X_{\em hep}}
{f_B\,\Phi_B^0\,\sigma_B^X}\,\rho_{\em hep}^X(E_\nu)}
{1+\frac{f_{\em hep}\,\Phi_{\em hep}^0\,\sigma^X_{\em hep}}
{f_B\,\Phi_B^0\,\sigma_B^X}}\ ,
\label{rhotot}
\end{eqnarray}
where our evaluations of the cross sections $\sigma^X_{\em hep}$ (defined in a
way analogous to $\sigma^X_{B}$) are given in the second, third, and fourth
column of Table~III for $X=e,a,c$, respectively.  In the same table, we also
report the ratios between the standard contributions of {\em hep\/} neutrinos
($\Phi^0_{\em hep}\,\sigma^X_{\em hep}$) to $^8$B neutrinos  ($\Phi^0_{\em
B}\,\sigma^X_{B}$) in each $i$-th energy range. Since such ratios are small,
the combined response functions $\rho^X$ [Eq.~(\ref{rhotot})] are always
dominated by the $\rho^X_B$ component, even for a relatively large {\em hep\/}
flux (e.g., $f_{\em hep}/f_B=3$). Therefore, we expect that, although the SK
and SNO response functions to {\em hep\/} neutrinos can be noticeably different
(see Fig.~4), the {\em combined\/} response functions to (B+{\em hep})
neutrinos  can still be taken as approximately equal in SK-SNO corresponding
ranges. Such expectation  is confirmed by Fig.~5, were the  combined response
functions are shown in the same representative energy ranges of Fig.~2, but for
the reference case of relatively large {\em hep\/} flux $(f_B,\,f_{\em
hep})=(1,\,3)$.%
\footnote{ One can appreciate the contribution of {\em hep\/} neutrinos in the
upper tail of the response functions  by comparing Fig.~5 ($f_{\em hep}=3$)
with Fig.~2 ($f_{\em hep}=0$).}
The SK and SNO combined response functions (solid and dotted curves,
respectively)  are indeed approximately equal in any energy range, the integral
difference $\Delta$ being a few percent (as reported in the last column of
Table~III), except for the last bin where it exceeds $10\%$. We remind that the
situation would be better (smaller $\Delta$) for the phenomenologically
preferred case \cite{Su00} of $f_{\em hep}/f_B<3$. Therefore, if we average the
oscillation probability over the {\em combined\/} response functions in each
$i$-th energy range,
\begin{equation}
\langle P_{ee}\rangle^{X,i} = \int dE_\nu\,\rho^{X,i}(E_\nu)
\,P_{ee}(E_\nu)\ ,
\end{equation}
we can still assume them as approximately equal to each other,
\begin{equation}
\langle P_{ee}\rangle^{i}\equiv
\langle P_{ee}\rangle^{e,i} =
\langle P_{ee}\rangle^{a,i} \simeq 
\langle P_{ee}\rangle^{c,i} \ \ (i=1,\dots,13)\ ,
\label{PEQUAL}
\end{equation}
as it was the case for $^8$B  neutrinos only [Eq.~(\ref{Pequal})]. Finally, we
remark  that the approximate equalization of the combined SK and SNO response
functions in Fig.~5 is partly due also to the fact that the relative {\em
hep\/} contribution in each energy range is approximately independent of the
interaction process $X(=e,a,c)$,
\begin{equation}
\epsilon_i \equiv 
\frac{\Phi^0_{\em hep}\,\sigma^e_{\em hep}}
{\Phi^0_B\,\sigma^e_B}=
\frac{\Phi^0_{\em hep}\,\sigma^a_{\em hep}}
{\Phi^0_B\,\sigma^a_B}\simeq 
\frac{\Phi^0_{\em hep}\,\sigma^c_{\em hep}}
{\Phi^0_B\,\sigma^c_B}\ ,
\label{eps}
\end{equation}
the differences being a few~$\times10^{-3}$  (compare the 5th, 6th, and 7th
columns of Table~III), except, once again, for the last range ($i=13$), where
the above approximation   is not as good.

Let us discuss the effect of a nonzero {\em hep\/} flux on the results of the
previous section. We redefine the expected normalized rates $r^i_{\rm SK}$ and
$r^i_{\rm SNO}$ as in Eqs.~(\ref{r_SK},\ref{r_SNO}), with denominators
$R^{0,i}_{\rm SK}$ and $R^{0,i}_{\rm SNO}$ still given by the standard rate
from $^8$B neutrinos {\em only\/},%
\footnote{The ``standard'' rate to which one should normalize the measured rate
may or may not include the SSM {\em hep\/}  neutrino contribution (in addition
to SSM $^8$B neutrinos),  the choice being purely conventional. Our choice
({\em hep\/} neutrinos not included in the standard unoscillated rates $R_{\rm
SK}^{0,i}$ and $R_{\rm SNO}^{0,i}$) leads to more compact expressions in the
context of our approach.}
as defined in Eqs.~(\ref{R0_SK}) and (\ref{R0_SNO}). However, we now include
the {\em hep\/} neutrino contribution in the numerators $R^{i}_{\rm SK}$ and
$R^{i}_{\rm SNO}$:
\begin{eqnarray}
R^{i}_{\rm SK}&=&  f_B\,\Phi^0_B\,
\left[\sigma_B^{e,i}\,\langle P_{ee}\rangle^{e,i}_B
+\sigma_B^{a,i}\,\left(1-\langle P_{ee}\rangle^{a,i}_B\right)\right]
\ +\ (B\leftrightarrow{\em hep})\ ,
\label{Rtot_SK}\\
R^{i}_{\rm SNO}&=&
f_B\,\Phi^0_B\,\sigma_B^{c,i}\,\langle P_{ee}\rangle^{e,i}_B
\ +\ (B\leftrightarrow{\em hep})\ .
\label{Rtot_SNO}
\end{eqnarray}

The above Eqs.~(\ref{Rtot_SK},\ref{Rtot_SNO}) are exact,  i.e., they do not
imply any approximation. By applying to such equations the approximate
equalities expressed by  Eqs.~(\ref{PEQUAL}) and (\ref{eps}) we obtain, after
some algebra,
\begin{eqnarray}
r^i_{\rm SK} & = &
(f_B+\epsilon_i f_{\em hep})\left[
 \langle P_{ee}\rangle^i + \frac{\sigma^{a,i}_B}{\sigma^{e,i}_B}
(1-\langle P_{ee}\rangle^i) \right]\ ,
\label{rtot_SK}\\
r^i_{\rm SNO} & = &
(f_B+\epsilon_i f_{\em hep})\,
 \langle P_{ee}\rangle^i \ ,
\label{rtot_SNO}
\end{eqnarray}
from which we can eliminate the average probability  $\langle P_{ee}\rangle^i$
to get
\begin{equation}
r^i_{\rm SK} = r^i_{\rm SNO} \left( 1-\frac{\sigma^{a,i}_B}{\sigma^{e,i}_B}
\right) + (f_B+\epsilon_i f_{\em hep})
\,\frac{\sigma^{a,i}_B}{\sigma^{e,i}_B}\ \ 
\ .
\label{MAIN}
\end{equation}

The above equation holds with an accuracy comparable to that of
Eq.~(\ref{main}), namely, $O(1\%)$ for typical solutions to the solar neutrino
problem. The only difference with Eq.~(\ref{main}) is  the replacement $f_B\to
f_B+\epsilon_if_{\em hep}$, where $\epsilon_i$ [defined in Eq.~(\ref{eps})] is
tabulated in the fifth column of Table~III. Therefore, all the considerations
in Sec.~IV, related to the existence of a linear relation between SK and SNO
rates in appropriate energy ranges, hold also in the presence of a
nonnegligible {\em hep\/} neutrino contribution, up to the mentioned
replacement for $f_B$ (barring perhaps the last range $i=13$, where the
approximations may not be particularly good).

In addition, suppose that both the $r^i_{\rm SK}$'s and the $r^i_{\rm SNO}$'s
are measured with high precision, and that $f_B$ is also precisely determined
from the NC measurement in SNO \cite{SNOB}: then  the only unknown  would be
$f_{\em hep}$, which could be hopefully determined by applying Eq.~(\ref{MAIN})
to the highest energy ranges (say, $i=11,\, 12$, and possibly $13$) where
$\epsilon_i$ is largest. Of course, at the level of accuracy required to
determine the absolute {\em hep\/} neutrino flux  $f_{\em hep}\Phi^0_{\em hep}$
from the SK and SNO data, one needs a very careful estimate of all the
uncertainties involved, including the intrinsic approximations of our
approach.

\section{Effect of Earth matter and of sterile neutrinos}

In this section we briefly discuss some limitations or modifications of our
results, which can arise when neutrino oscillations are affected by Earth
matter effects \cite{Wo78,Mi85,Ba80},  or by transitions to a sterile state
$\nu_s$.

\subsection{Earth matter effects}

So far, we have implicitly assumed one and the same probability function
$P_{ee}(E_\nu)$ for both SK and SNO. This assumption is no longer valid in the
presence of sizable Earth matter effects (during nighttime), since the
different SK and SNO latitudes imply different nadir angle ($\eta$)  exposures
for the two experiments \cite{Li97}. Although the SK data have already excluded
a significant fraction of the parameter space where Earth matter effects are
large \cite{Su00},  one cannot exclude that more precise SK (and SNO) data may
still reveal some distortion in the nadir distribution  (see, e.g., 
\cite{CCer,Zeni} and references therein). In such case, the probability
function $P_{ee}$ could be significantly different in SK and SNO, especially
along the inner trajectories  traversing the Earth core, where mantle+core
interference effects can occur \cite{Core},  and where the exposure functions
in SK and SNO are significantly different.

It follows that, if sizable  Earth matter effects would emerge in SK or SNO
nighttime data,  our approach could be strictly applied only to daytime events.
Notice, however, that the SK and SNO exposure functions, although different in
principle, turn out to be approximately similar in the nadir angle range
$0.4\lesssim \eta \lesssim 0.8$ (see Fig.~3 in \cite{Li97}). Therefore, modulo
such approximation, one could still assume one and the same function
$P_{ee}(E_\nu)$ in SK and SNO  (and thus apply our results) for the fraction of
nighttime data with $\eta\in[0.4,\,0.8]$.

\subsection{Sterile neutrino oscillations}

Recent global fits to solar neutrino data, including the latest SK results
\cite{Su00}, tend to disfavor pure  $\nu_e\to\nu_s$ oscillations
\cite{Su00,Four} as compared to $\nu_e\to\nu_{\mu,\tau}$, although it is
perhaps too early to claim rejection of the $\nu_s$ scenario  \cite{Four}. In
any case, mixed (active+sterile) solar neutrino oscillations are certainly
still allowed \cite{Four}, and can also be made consistent with the atmospheric
neutrino oscillation evidence \cite{Marr}.  In the general case of
active+sterile oscillations,  Eqs.~(\ref{r_SK}) and (\ref{r_SNO}) are modified
as follows: 
\begin{eqnarray}
r^i_{\rm SK} & = &
f_B\,\left[
 \langle P_{ee}\rangle^i + \frac{\sigma^{a,i}_B}{\sigma^{e,i}_B}
(1-\langle P_{ee}\rangle^i-\langle P_{es}\rangle^i) \right]\ ,
\label{rster_SK}\\
r^i_{\rm SNO} & = &
f_B\,
 \langle P_{ee}\rangle^i \ ,
\label{rster_SNO}
\end{eqnarray}
where the unitarity relation $1=P_{ee}+P_{e\mu}+P_{es}$ has been used, $P_{es}$
being the $\nu_e\to\nu_s$ oscillation probability.%
\footnote{Similar relations were discussed in  \protect\cite{Vill} for the
total SK and SNO rates.}

Unfortunately, one cannot eliminate both $\langle P_{ee}\rangle^i$  and
$\langle P_{es}\rangle^i$  from the above equations, so as to derive a
model-independent relation between the SK and SNO rates analogous to
Eq.~(\ref{main}).  However, one can eliminate at least $\langle
P_{ee}\rangle^i$ to get   the relation
\begin{equation}
\frac{r^i_{\rm SK}-r^i_{\rm SNO}(1-\sigma_B^{a,i}/\sigma_B^{e,i})}
{\sigma_B^{a,i}/\sigma_B^{e,i}}
= f_B\left(1-\langle P_{es}\rangle^i\right)\ ,
\label{rela}
\end{equation}
which might still provide a useful test of $\nu_e\to\nu_s$ transitions. In
fact, the quantity $f_B\left(1-\langle P_{es}\rangle^i\right)$ is independent
of energy range index $i$ only if $P_{es}(E_\nu)=\rm const$. Therefore,
variations of the left-hand-side of Eq.~(\ref{rela}) in different energy ranges
would provide a signature of $\nu_e\to\nu_s$ transitions  with energy-dependent
probability.

Further links between $r^i_{\rm SK}$ and $r^i_{\rm SNO}$ can only be obtained
by making some assumptions about $P_{es}$. For instance, by assuming pure
sterile oscillations ($1=P_{ee}+P_{es}$), one gets from Eqs~(\ref{rster_SK})
and (\ref{rster_SNO}) that:
\begin{equation}
{\rm pure\ }\nu_e\to\nu_s\ {\rm oscillations}
\Longrightarrow
r^i_{\rm SK}=r^i_{\rm SNO} \ ,
\label{mains}
\end{equation}
i.e., the normalized rates in the SK and SNO corresponding ranges are expected
to be equal (due to the absence of $\nu_{\mu,\tau}$ NC contributions in SK).
The above equation is interesting because, independently of the functional form
of $P_{ee}(E_\nu)$,  it provides a possible way to distinguish pure sterile
oscillations  [Eq.~(\ref{mains})] from pure active oscillations
[Eq.~(\ref{main})]. Figure~6 shows, for instance, the ``predicted'' SNO
spectrum  for pure $\nu_e\to\nu_s$ oscillations, as obtained from
Eq.~(\ref{mains}). A comparison with Fig.~3 (``predicted'' SNO spectrum for
active oscillations) shows how the normalized SNO rates are expected to be less
suppressed for the sterile case, on the basis of the present SK data. Finally,
notice that Eq.~(\ref{mains}) does not allow to determine $f_B$, nor to
discriminate pure sterile oscillations from no oscillations. Such  loss of
predictive power  simply reflects the fact that, for increasing amplitude of
the $\nu_e\to\nu_s$ channel, there is a decreasing NC contribution in SK, so
that the SK and SNO event rates tend to be equally suppressed, their
combination providing eventually little additional information.

\section{Conclusions}

In this work, we have shown that the SK detector response to $^8$B solar
neutrinos, in each bin of the electron energy spectrum above 8 MeV, can be
accurately approximated by the  SNO detector response in appropriate
(different) electron energy ranges. In a sense, it is possible to ``tune'' the
SNO energy ``bandwidth'' so as to ``equalize'' the SK response in various
spectrum bins. As a consequence, we have derived a set of linear relations
among the SK and SNO spectral rates, whose distinguishing feature is the
independence on the functional form of the  (active) neutrino oscillation
probability.  Such relations can be used to determine the {\em absolute\/}
$^8$B neutrino flux from SK and SNO data,  and to cross-check the joint
(non)observation of spectral deviations in the two experiments. As an exercise,
pending official SNO spectrum data, we have inverted such relations to
``predict'' the SNO energy spectrum,  on the basis of the current SK data. We
have also discussed the effects and implications of a nonnegligible {\em hep\/}
neutrino flux, of  relatively large Earth matter effects, and of sterile
neutrino transitions.  We conclude by stressing that, should our results be
adopted by the SK and SNO collaboration to perform a joint SK-SNO spectral
analysis, dedicated MonteCarlo simulations would be desirable to assess all the
uncertainties and their correlations, as well as to quantify (through a more
detailed description of the instrumental responses) the approximations
associated with our approach.

\acknowledgments

This work was supported in part by INFN (FA51 Project)  and in part by the
Italian MURST (``Astroparticle Physics'' Project). We thank D.\ Montanino for
useful discussions and suggestions.


\begin{table}
\caption{SNO electron energy ranges (third column) where the SNO response
function to $^8$B neutrinos equalizes the SK response function in the $i$-th
electron energy bin (second column). The integral difference $\Delta$ between
the SK and SNO response functions is given in the fourth column. $^8$B neutrino
cross sections for electron production in each range (including detector
resolution effects) are given in the remaining columns. See the text for
details.}
\begin{tabular}{cccccccc}
Range  						& 
$[E_e^{\min},\,E_e^{\max}]$ 			&
$[\tilde{E}_e^{\min},\,\tilde{E}_e^{\max}]$ 	& 
$\Delta$					&
$\sigma_B^e$    				&
$\sigma_B^a$    				&
$\sigma_B^a/\sigma_B^e$ 	   		&
$\sigma_B^c$\\
$i$						&
(SK, MeV)					&
(SNO, MeV)					&
$\times 100$					&
($10^{-44}$ cm$^2$)				&
($10^{-45}$ cm$^2$)				&
						&
($10^{-42}$ cm$^2$)				\\
\hline
 1&$[8,\,8.5]$  &$[5.10,\,9.90]$  &6.9&0.1415&0.2191&0.155&0.8092 \\
 2&$[8.5,\,9]$  &$[5.60,\,10.35]$ &5.2&0.1182&0.1819&0.154&0.7740 \\
 3&$[9,\,9.5]$  &$[6.10,\,10.80]$ &3.8&0.0970&0.1485&0.154&0.7214 \\
 4&$[9.5,\,10] $&$[6.60,\,11.30]$ &2.8&0.0781&0.1190&0.153&0.6570 \\
 5&$[10,\,10.5]$&$[7.10,\,11.75]$ &2.1&0.0616&0.0934&0.152&0.5790 \\
 6&$[10.5,\,11]$&$[7.55,\,12.30]$ &1.8&0.0475&0.0718&0.151&0.5070 \\
 7&$[11,\,11.5]$&$[8.05,\,12.85]$ &1.8&0.0357&0.0538&0.151&0.4222 \\
 8&$[11.5,\,12]$&$[8.50,\,13.45]$ &2.0&0.0262&0.0394&0.150&0.3481 \\
 9&$[12,\,12.5]$&$[8.95,\,14.35]$ &2.6&0.0186&0.0280&0.150&0.2786 \\
10&$[12.5,\,13]$&$[9.45,\,14.95]$ &3.1&0.0129&0.0193&0.150&0.2087 \\
11&$[13,\,13.5]$&$[9.90,\,18.25]$ &3.8&0.0086&0.0129&0.150&0.1554 \\
12&$[13.5,\,14]$&$[10.30,\,20]$   &4.4&0.0056&0.0083&0.148&0.1159 \\
13&$[14,\,20]  $&$[11.20,\,20]$   &8.6&0.0081&0.0120&0.148&0.0532 
\end{tabular}
\end{table}

\begin{table}
\caption{Linear relations linking the normalized SK and SNO spectral rates
(last column) in SK-SNO corresponding energy ranges (reported for completeness
in the 2nd and 3rd columns). Such relations hold for any functional form on the
$\nu_e$ survival probability, assuming active $\nu$ oscillations.}
\begin{tabular}{cccl}
Range  						& 
$[E_e^{\min},\,E_e^{\max}]$ 			&
$[\tilde{E}_e^{\min},\,\tilde{E}_e^{\max}]$ 	& 
Relation between normalized 		\\
$i$						&
(SK, MeV)					&
(SNO, MeV)					&
event rates in SK and SNO			\\
\hline
 1&$[8,\,8.5]$  &$[5.10,\,9.90]$  &$r^{}_{\rm SK}=0.845\times r^{}_{\rm SNO}+0.155\times f_B$ \\
 2&$[8.5,\,9]$  &$[5.60,\,10.35]$ &$r^{}_{\rm SK}=0.846\times r^{}_{\rm SNO}+0.154\times f_B$ \\
 3&$[9,\,9.5]$  &$[6.10,\,10.80]$ &$r^{}_{\rm SK}=0.846\times r^{}_{\rm SNO}+0.154\times f_B$ \\
 4&$[9.5,\,10] $&$[6.60,\,11.30]$ &$r^{}_{\rm SK}=0.847\times r^{}_{\rm SNO}+0.153\times f_B$ \\
 5&$[10,\,10.5]$&$[7.10,\,11.75]$ &$r^{}_{\rm SK}=0.848\times r^{}_{\rm SNO}+0.152\times f_B$ \\
 6&$[10.5,\,11]$&$[7.55,\,12.30]$ &$r^{}_{\rm SK}=0.849\times r^{}_{\rm SNO}+0.151\times f_B$ \\
 7&$[11,\,11.5]$&$[8.05,\,12.85]$ &$r^{}_{\rm SK}=0.849\times r^{}_{\rm SNO}+0.151\times f_B$ \\
 8&$[11.5,\,12]$&$[8.50,\,13.45]$ &$r^{}_{\rm SK}=0.850\times r^{}_{\rm SNO}+0.150\times f_B$ \\
 9&$[12,\,12.5]$&$[8.95,\,14.35]$ &$r^{}_{\rm SK}=0.850\times r^{}_{\rm SNO}+0.150\times f_B$ \\
10&$[12.5,\,13]$&$[9.45,\,14.95]$ &$r^{}_{\rm SK}=0.850\times r^{}_{\rm SNO}+0.150\times f_B$ \\
11&$[13,\,13.5]$&$[9.90,\,18.25]$ &$r^{}_{\rm SK}=0.850\times r^{}_{\rm SNO}+0.150\times f_B$ \\
12&$[13.5,\,14]$&$[10.30,\,20]$   &$r^{}_{\rm SK}=0.852\times r^{}_{\rm SNO}+0.148\times f_B$ \\
13&$[14,\,20]  $&$[11.20,\,20]$   &$r^{}_{\rm SK}=0.852\times r^{}_{\rm SNO}+0.148\times f_B$
\end{tabular}
\end{table}

\begin{table}
\caption{Cross sections for electron production by {\em hep\/} neutrinos in the
SK and SNO corresponding energy ranges. The ratio of  {\em hep\/} to $^8$B
contributions in each bin is also given, assuming standard BP 2000 neutrino
fluxes: $\Phi_{B}^{\rm 0}=5.15\times 10^6$ cm$^{-2}$~s$^{-1}$ and $\Phi_{\it
hep}^{0}=9.3\times 10^3$ cm$^{-2}$~s$^{-1}$ \protect\cite{BP00}. The last
column gives the integral difference among the SK and SNO response functions
for the reference case of ``large {\em hep\/} flux'' ($f_{\em hep}=3$ and
$f_B=1$), as discussed in the text.}
\begin{tabular}{cccccccc}
Range  						& 
$\sigma_{\it hep}^e$	 			&
$\sigma_{\it hep}^a$			 	& 
$\sigma_{\it hep}^c$				&
$\sigma_{\it hep}^e/\sigma_B^e$   		&
$\sigma_{\it hep}^a/\sigma_B^a$    		&
$\sigma_{\it hep}^c/\sigma_B^c$   		&
$\Delta(f_{\it hep}=3)$					\\[1mm]
$i$						&
($10^{-44}$ cm$^2$)				&
($10^{-45}$ cm$^2$)				&
($10^{-42}$ cm$^2$)				&
$\times\Phi_{\sl hep}^{0}/\Phi_{B}^{0}$ 	&
$\times\Phi_{\sl hep}^{0}/\Phi_{B}^{0}$   	&
$\times\Phi_{\sl hep}^{0}/\Phi_{B}^{0}$ 	&
$\times 100$					\\[1mm]
\hline
 1&0.2826&0.4476&1.2478&0.004&0.004&0.003&7.2 \\
 2&0.2615&0.4107&1.3449&0.004&0.004&0.003&5.4 \\
 3&0.2403&0.3745&1.4285&0.005&0.005&0.004&4.1 \\
 4&0.2192&0.3391&1.5137&0.005&0.005&0.004&3.1 \\
 5&0.1982&0.3048&1.5605&0.006&0.006&0.005&2.5 \\
 6&0.1777&0.2718&1.6317&0.007&0.007&0.006&2.1 \\
 7&0.1579&0.2402&1.6613&0.008&0.008&0.007&2.2 \\
 8&0.1388&0.2103&1.6881&0.010&0.010&0.009&2.3 \\
 9&0.1207&0.1821&1.7395&0.012&0.012&0.011&2.7 \\
10&0.1037&0.1560&1.6646&0.015&0.015&0.014&3.1 \\
11&0.0880&0.1319&1.6893&0.018&0.018&0.020&3.8 \\
12&0.0736&0.1100&1.5465&0.024&0.024&0.024&4.4 \\
13&0.2576&0.3823&1.2133&0.057&0.057&0.041&13.2
\end{tabular}
\end{table}

\newcommand{\InsertFigure}[2]{\newpage\phantom{.}
\vspace*{-2.cm}\begin{center}\mbox{%
\epsfig{bbllx=2truecm,bblly=2truecm,bburx=19.5truecm,bbury=26.5truecm,%
height=23.truecm,figure=#1}}\end{center}\vspace*{-2.85truecm}%
\parbox[t]{\hsize}{\small\baselineskip=0.5truecm\hskip0.5truecm #2}}
\newcommand{\LowerFigure}[2]{\newpage\phantom{.}
\vspace*{-1.6cm}\begin{center}\mbox{%
\epsfig{bbllx=2truecm,bblly=2truecm,bburx=19.5truecm,bbury=26.5truecm,%
height=23.truecm,figure=#1}}\end{center}\vspace*{-2.9truecm}%
\parbox[t]{\hsize}{\small\baselineskip=0.5truecm\hskip0.5truecm #2}}

\InsertFigure{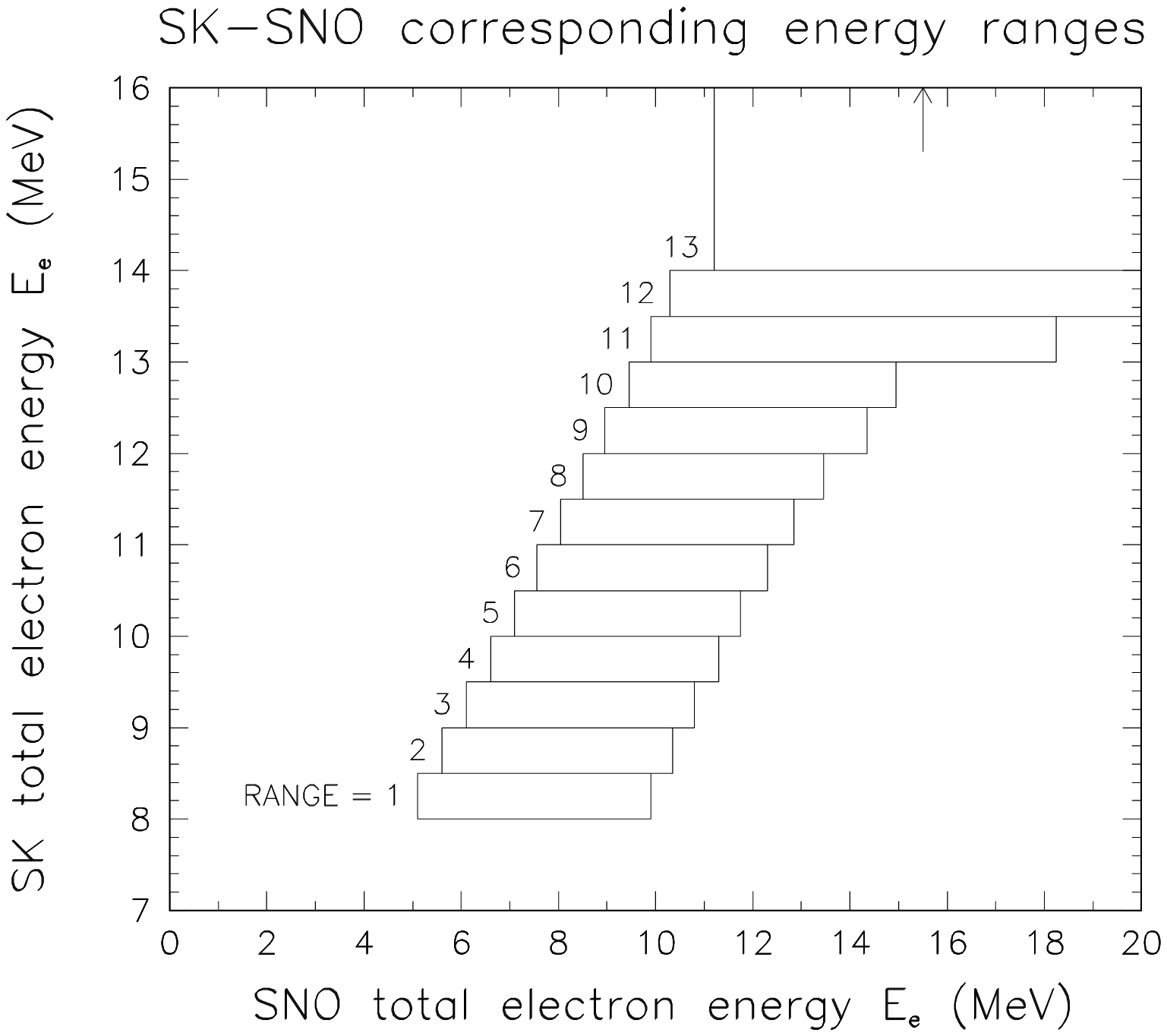}%
{FIG.~1. Correspondence between electron energy ranges in SK and SNO. For each
SK energy bin on the $y$-axis, the associated energy range in SNO  is given on
the $x$-axis. Ranges are labelled by sequential numbers. See also Table~I. 
}
\InsertFigure{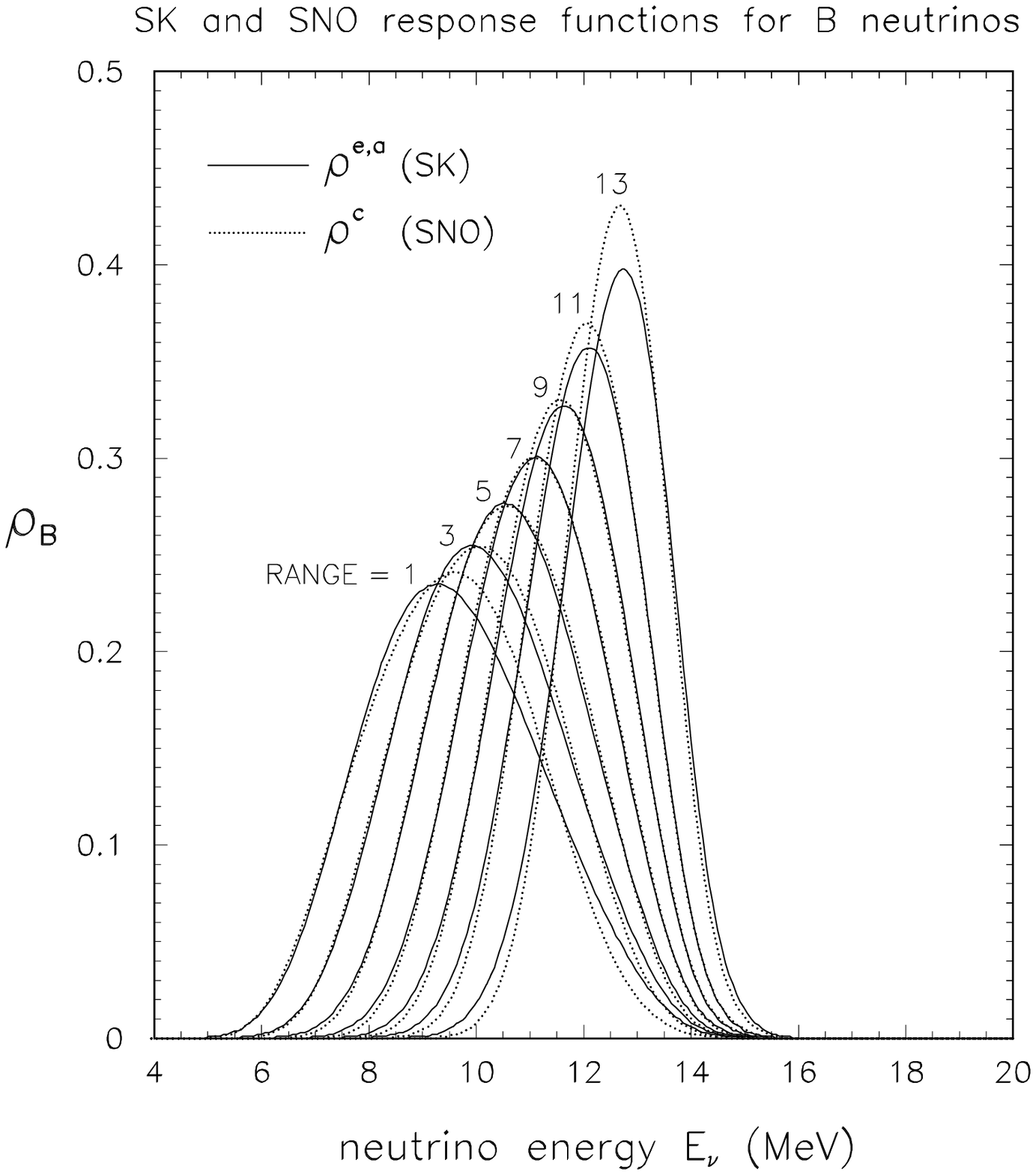}%
{FIG.~2. SK and SNO response functions to $^8$B neutrinos in a representative
set of energy ranges. The functions are nearly coincident (``equalized'')
within a few percent.
}
\LowerFigure{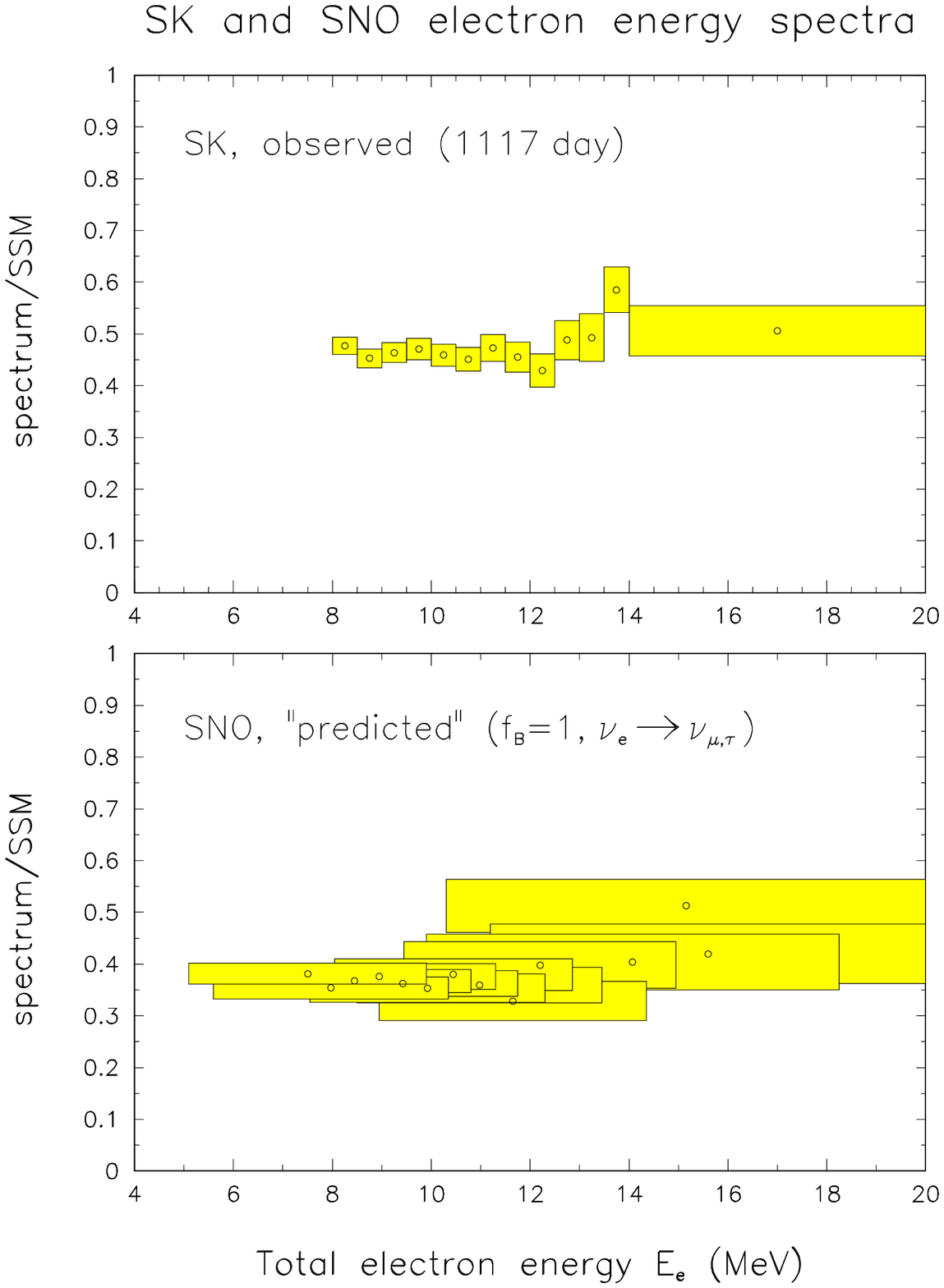}%
{FIG.~3. Upper panel: SK observed energy spectrum  (1117 day exposure
\protect\cite{Su00}), normalized to the  BP 2000 expectations for $^8$B
neutrinos \protect\cite{BP00}. The width and height  of each grey box represent
the bin energy range and the $\pm1\sigma$  statistical  uncertainty,
respectively. Lower panel: SNO ``predicted'' energy spectrum, as obtained by
projecting each SK energy bin rate onto  the SNO correlated energy range (and
propagating the SK statistical errors). The ``prediction'' (made for $f_B=1$)
is valid for  active oscillations, independently of the functional form of the
flavor transition probability. See the text for details.
}
\InsertFigure{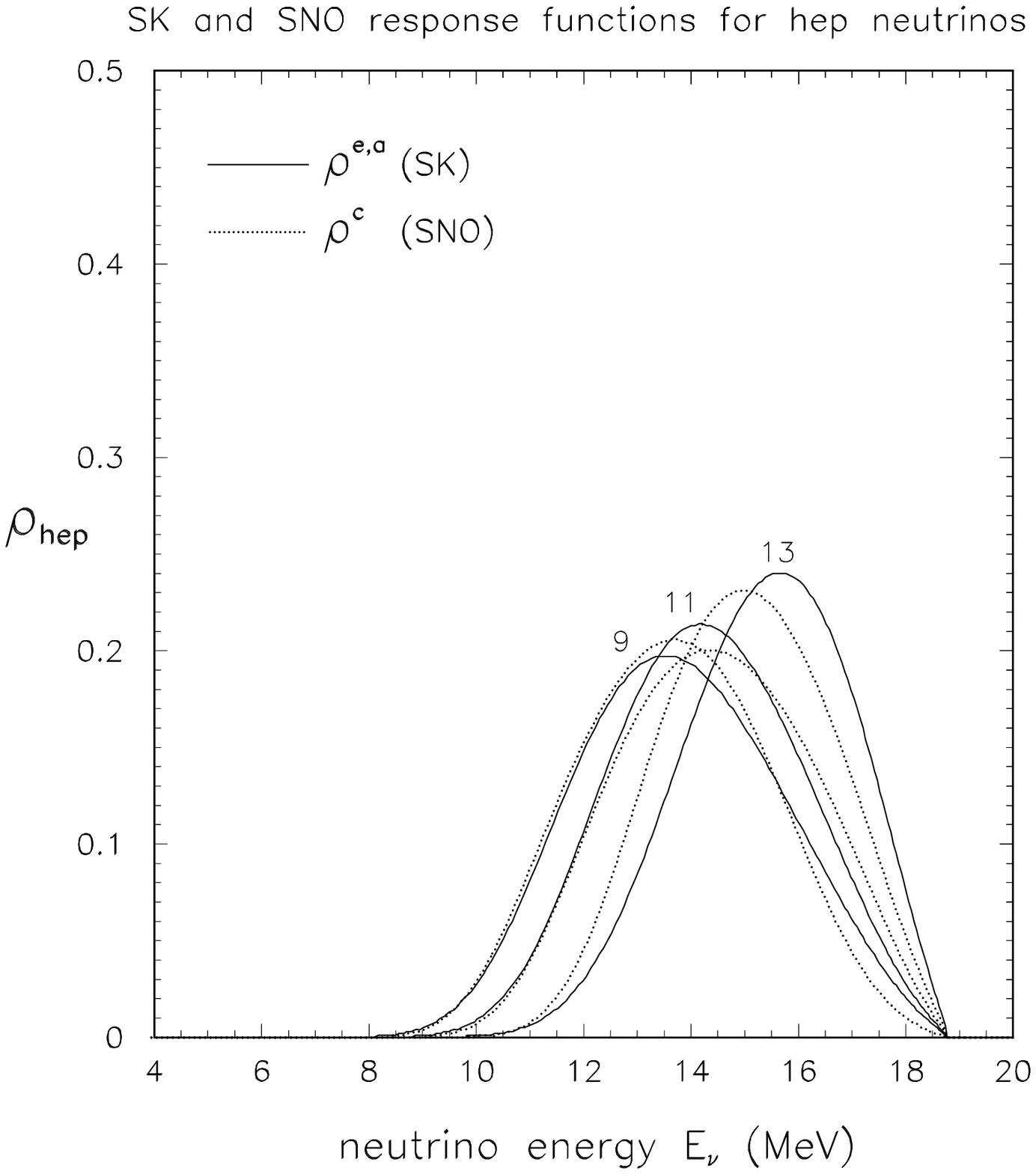}%
{FIG.~4. SK and SNO response functions to {\em hep\/} neutrinos for three
representative (high) energy ranges. 
}
\InsertFigure{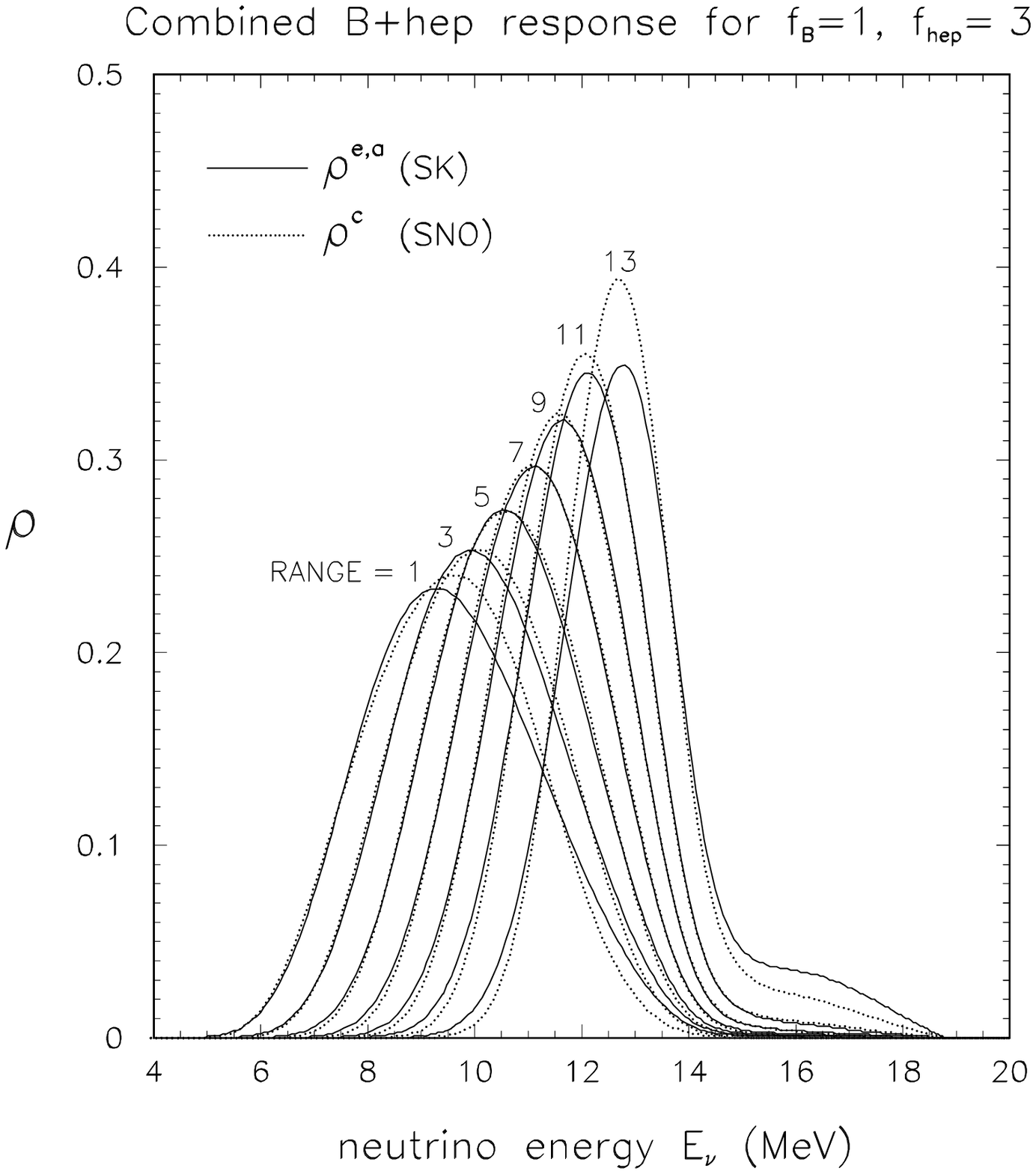}%
{FIG.~5. Combined SK and SNO response functions to $^8$B and {\em hep\/}\
neutrinos, assuming a {\em hep\/} neutrino flux three times larger than the
BP~2000 expectations \protect\cite{BP00}. The functions are approximately
coincident (typically within a few percent). A comparison with Fig.~2 shows
that the {\em hep\/} contribution modifies the upper tail of the curves  for
the highest energy ranges, but does not really spoil the approximate equality
between the SK and SNO response functions. 
}
\InsertFigure{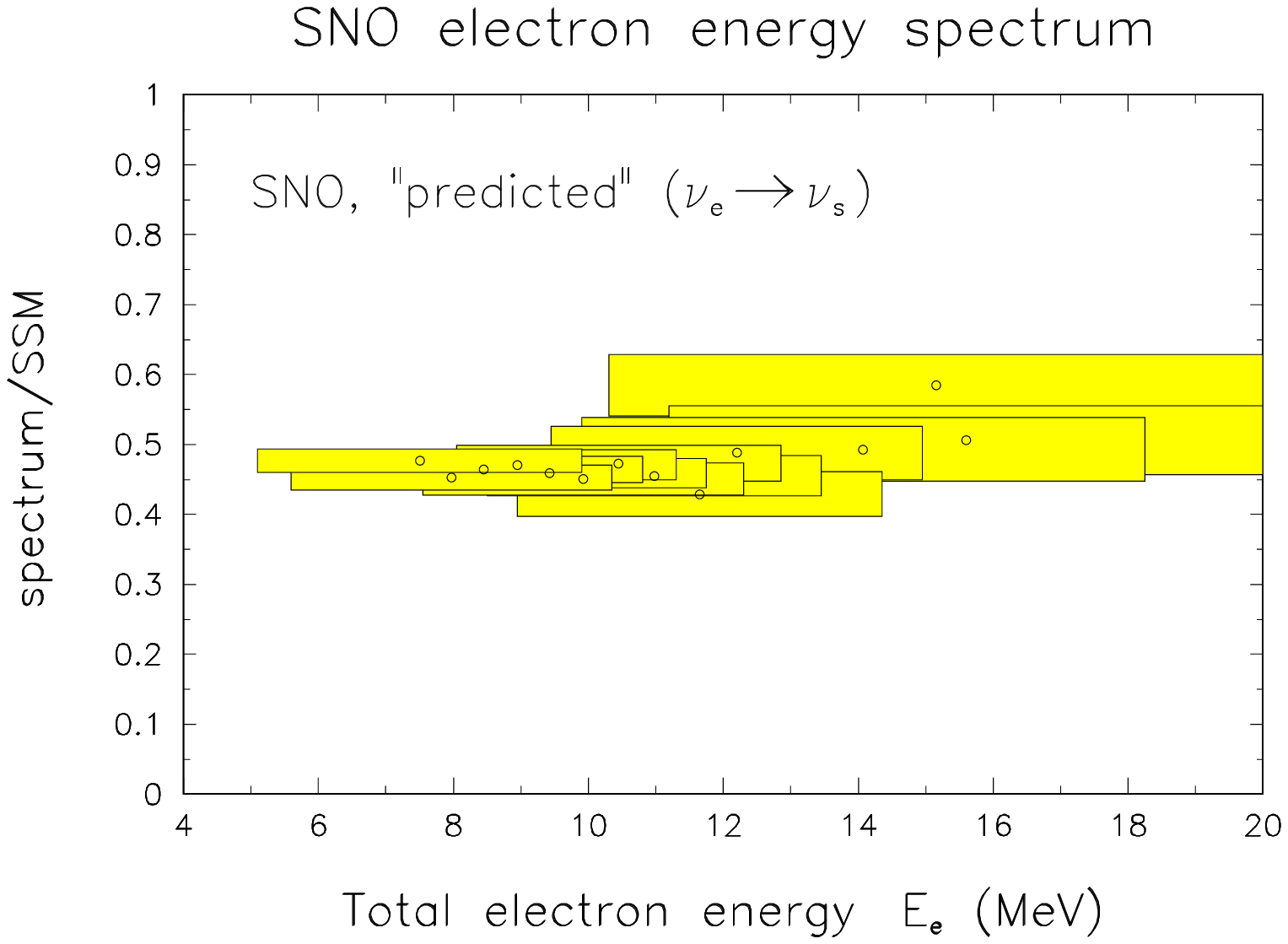}%
{FIG.~6. As in the lower panel in Fig.~3, but for the case of sterile neutrino
oscillations.
}



\begin{thebibliography}{99}

\bibitem{Cl98}	Homestake Collaboration, 
		B.T.\ Cleveland, T.J.\ Daily, R.\ Davis Jr., J.R.\
		Distel, K.\ Lande, C.\ K.\ Lee, P.S.\ Wildenhain, and J.\
		Ullman, 
		Astrophys.\ J.\ {\bf 496}, 505 (1998).

\bibitem{Fu96}	Kamiokande Collaboration, 
		Y.\ Fukuda {\em et al.},
		Phys.\ Rev.\ Lett.\ {\bf 77}, 1683 (1996).

\bibitem{Ab99}	SAGE Collaboration, 
		J.N.\ Abdurashitov {\em et al.},
		Phys.\ Rev.\ C {\bf 60}, 055801 (1999).

\bibitem{Ha99}	GALLEX Collaboration, 
		W.\ Hampel {\em et al.},
		Phys.\ Lett.\ B {\bf 447}, 127 (1999).

\bibitem{GNOC}	GNO Collaboration,
		M.\ Altmann {\em et al.},
		Phys.\ Lett.\ B {\bf 490}, 16 (2000).

\bibitem{SK00}	Super-Kamiokande Collaboration, 
		Y.\ Fukuda {\em et al.},
		Phys.\ Rev.\ Lett.\  {\bf 82}, 2430 (1999).

\bibitem{NuAs}	J.\ N.\ Bahcall, 
		{\em Neutrino Astrophysics} (Cambridge University Press,
		Cambridge, England, 1989).

\bibitem{BP00}	J.N.\ Bahcall, M.H.\ Pinsonneault, and S.\ Basu,
		astro-ph/0010346.

\bibitem{JNBH}	J.N.\ Bahcall homepage, http://www.sns.ias.edu/$^\sim$jnb\ 
		(Neutrino Software and Data).

\bibitem{Su00}	Y.\ Suzuki for the Super-Kamiokande Collaboration, 
		in {\em Neutrino~2000},
		19th International Conference on Neutrino
		Physics and Astrophysics (Sudbury, Canada, 2000)
		Nucl.\ Phys.\ B (Proc.\ Suppl.) 91, 29 (2001).

\bibitem{SNOB}	A.B.\ McDonald for the SNO Collaboration, 
		in {\em Neutrino~2000} \protect\cite{Su00}, p.~21.

\bibitem{BaLi}	J.N.\ Bahcall and E.\ Lisi,
		Phys.\ Rev.\ D {\bf 54}, 5417 (1996).

\bibitem{Faid}	B.\ Fa{\"\i}d, G.L.\ Fogli, E.\ Lisi, and D.\ Montanino,
		Phys.\ Rev.\ D {\bf 55}, 1353 (1997).
		
\bibitem{300d}	G.L.\ Fogli, E.\ Lisi, and D.\ Montanino,
		Astropart.\ Phys.\ {\bf 9}, 119 (1998).

\bibitem{Vill}	F.L.\ Villante, G.\ Fiorentini, and E.\ Lisi,
		Phys.\ Rev.\ D {\bf 59}, 013006 (1999).

\bibitem{Bspe}	J.N.\ Bahcall, E.\ Lisi, D.E.\ Alburger,
		L.\ De Braeckeleer, S.J.\ Freedman, and J.\ Napolitano,
		Phys.\ Rev.\ C {\bf 54}, 411 (1996).

\bibitem{EScs}	We adopt the $\nu$-$e$ cross section calculations 
		reported in J.N.\ Bahcall, M.\ Kamionkowsky, and A.\ Sirlin,
		Phys.\ Rev.\ D {\bf 51}, 6146 (1995).
		See also M.\ Passera, hep-ph/0011190, for recent refined
		calculations of $O(\alpha)$ QED corrections to $\nu$-$e$
		scattering.
	
\bibitem{CCcs}	We adopt the $\nu$-$d$ CC cross section
		as reported in the computer code {\tt DIFFCROSS.FOR\/}
		\protect\cite{JNBH,BaLi}, and corresponding to the
		``Kubodera-Nozawa'' calculation option. See also
		S.~Nakamura, T.\ Sato, V.\ Gudkov, and K.\ Kubodera,
		nucl-th/0009012, for recent refined calculations of
		$\nu$-$d$ cross sections.
		
\bibitem{SKre}	In the present work, the width of the SK energy resolution
		function is taken to be $\pm 15\%$ for an electron (kinetic)
		energy 10 MeV, as in \protect\cite{300d}. Such value, which
		is in agreement with the SK calibration results
		graphically reported in \protect\cite{SK00}, scales
		with energy as described in \cite{BaLi,Faid}.
		
\bibitem{SNre}	In the present work, the width of the SNO energy resolution
		function is taken to be $\pm 11\%$ for an electron (kinetic)
		energy of 10 MeV, as in \protect\cite{BaLi}. Such value is in
		agreement with the preliminary SNO electron energy
		calibration results ($\sim9$ photomultiplier hits/MeV,
		as reported in 
		\protect\cite{SNOB}). The value previously adopted in 
		\protect\cite{Vill} (on the basis of the SNO prospective
		estimates available at that time) 
		was somewhat larger $(\pm 14\%)$.

\bibitem{Hint}	See., e.g., the SK-SNO spectral correlation studies in:
		S.M.\ Bilenky and C.\ Giunti, 
		Astropart.\ Phys.\ {\bf 2}, 353 (1994);
		K.\ Kwong amd S.P.\ Rosen,
		Phys.\ Rev.\ D {\bf 51}, 6159 (1995);
		G.L.\ Fogli, E.\ Lisi, and D.\ Montanino, 
		Phys.\ Lett.\ B {\bf 434}, 333 (1998);
		J.N.\ Bahcall, P.I.\ Krastev, and A.Yu.\ Smirnov,
		Phys.\ Rev.\ D {\bf 63}, 053012 (2001).

\bibitem{SKSP}	We use the SK spectrum data corresponding to 1117 days of 
		detector exposure \protect\cite{Su00}, normalized to the 
		$^8$B neutrino flux from the BP 2000 standard solar model 
		\protect\cite{BP00}.

\bibitem{CCer}	J.N.\ Bahcall, P.I.\ Krastev, and A.Yu.\ Smirnov,
		Phys.\ Rev.\ D {\bf 62}, 093004 (2000).

\bibitem{Wo78}	L.\ Wolfenstein, 
		Phys.\ Rev.\ D {\bf 17}, 2369 (1978).

\bibitem{Mi85}	S.P.\ Mikheyev and A.Yu.\ Smirnov, 
		Yad.\ Fiz.\ {\bf 42}, 1441 (1985) 
		[Sov.\ J.\ Nucl.\ Phys.\ {\bf 42}, 913 (1985)];
		Nuovo Cimento C {\bf 9}, 17 (1986).

\bibitem{Ba80}  V.\ Barger, S.\ Pakvasa, R.J.N.\ Phillips and K.\ Whisnant,
		Phys.\ Rev.\ D {\bf 22}, 2718 (1980).

\bibitem{Li97}	E.\ Lisi and D.\ Montanino, 
		Phys.\ Rev.\ D {\bf 56}, 1792 (1997). 

\bibitem{Zeni}	M.C.\ Gonzalez-Garcia, C.\ Pe{\~n}a-Garay, and A.Yu.\ Smirnov,
		hep-ph/0012313.

\bibitem{Core}	S.T.\ Petcov, 
		Phys.\ Lett.\ B {\bf 434}, 321 (1998); 
		E.Kh.\ Akhmedov, 
		Nucl.\ Phys.\ B {\bf 538}, 25 (1999).

\bibitem{Four}	C.\ Giunti, M.C.\ Gonzalez-Garcia, and C.\ Pe{\~n}a-Garay, 
		Phys.\ Rev.\ D {\bf 62}, 013005 (2000);
		M.C.\ Gonzalez-Garcia and C.\ Pe{\~n}a-Garay,
		hep-ph/0011245.
		
\bibitem{Marr}	G.L.\ Fogli, E.\ Lisi, and A.\ Marrone,
		Phys.\ Rev.\ D {\bf 63}, 053008 (2001).
		

\end{thebibliography}
\end{document}